# Calculation of the Sun exposure time for the synthesis of vitamin D in Urcuquí, Ecuador

Salum GM[1], García Molleja J[1], Regalado Díaz BA[1], Guerrero León LA[1] and Berrezueta C[1]

Yachay Tech University, School of Physical Sciences and Nanotechnology, 100119-Urcuquí, Ecuador
gsalum@yachaytech.edu.ec, jgarcia@yachaytech.edu.ec, bregalado@yachaytech.edu.ec, lenin.guerrero@yachaytech.edu.ec, luisa.berrezueta@yachaytech.edu.ec

**Abstract.** The synthesis of vitamin D is strongly linked to the availability of solar energy. For a long time, it was clear that using the erythemal irradiances is a good choice to calculate the exposure time. Several authors argue that this method of minimum exposure time overestimates the calculation of the solar irradiance of vitamin D. In this paper, the vitamin D and erythemal irradiances are calculated on the basis of the spectral solar irradiance in Urcuquí (Ecuador). From these results, we obtain the minimal exposure time. It was found that there is a difference between the use of the erythemal irradiance and the vitamin D one; namely, the exposure times were too large (almost a 100% difference) for the erythemal irradiance.

## 1 Introduction

Human beings do not produce the required daily dose of vitamin D (calciferol) on their own, other mechanisms must be used to obtain this minimum dose, *i.e.*, exposure to sunlight or dietary supplements [1]. Vitamin D is directly involved in the body's Ca-P homeostasis and its deficiency can cause rickets in children and osteomalacia in adults' homeostasis. Similarly, other important physiological processes linked to the immune system have been attributed to vitamin D deficiency. 25(OH) levels of vitamin D in the human body maintain healthy bones and the integrity of the immune and muscular systems [2]. Studies related to the emergence of erythema due to Sun exposure have demonstrated that short exposure times promote vitamin D synthesis but, on the other hand, long exposure times can promote skin cancer, cataracts and the suppression of the immune system [3]. It is worth mentioning that *erythema* is the reddening of the skin due to Sun exposure and for each skin phototype there is a minimum dose of ultraviolet energy that produces erythema (*minimal erythema dose* or *MED*). This MED depends on the irradiated skin zone and the number of exposures [4].
In 1975, Thomas Fitzpatrick changed the classification of the skin based on the color and responses to sun exporsure in terms of degrees of burning and tanning. This classification comprises six different skin types, from very fair (phototype "I") to very dark (phototype "VI") [5].
Some of the physical factors that have significant influence on the solar radiation are: altitude above sea level, albedo (reflectivity) of the ground, amount and types of aerosols (any solid or liquid particles suspended in the atmosphere) and the total amount of atmospheric ozone. The latter has an important influence on the level of UVB radiation [6]. If the altitude above sea level is increased, the total and UV (ultraviolet) solar irradiance increase, too. In particular, there is a mean increase of the global radiation ≈ 9% for every 1000 m of altitude and 23.7% in the erythema UV solar radiation [7]. In particular, it was found that in the Andean mountains the effect of altitude for UVB irradiance was only 7 – 9%, because in this region the amount of aerosols and tropospheric ozone is very small [8]. Then, it is worth noting that the levels of erythemal and vitamin D solar irradiance in Urcuquí are different from the ones in cities on the coast, at sea level.
For example, with the satellite database of the OMI/Aura satellite on the Giovanni/NASA webpage we extracted the daily average UV Index (erythemal solar irradiance multiplied by 40 [$m^2$/W]) in Urcuquí and Muisne (Ecuadorian city approximately at the same latitude than Urcuquí but at 100 masl) for the year 2014: $IUV_{Urcuqui}$ = 11.7 and $IUV_{Muisne}$ = 9.7. Thus, if we analyze the erythemal solar irradiance we obtain an erythemal solar irradiance of 0.2925 [W/$m^2$] for Urcuquí and 0.2425 [W/$m^2$] for Muisne which is equivalent to a difference of 20.2% and a positive variation of 9.4% per 1000 m of altitude increase.
The Institute of Medicine (US) has recommendations for the adequate daily intake of vitamin D according to age range, but most experts agree that without adequate Sun exposure, children and adults require approximately 800 to 1000 IU per day [9]. *Holick's rule* says that ¼ MED generates enough energy to produce vitamin $D_3$, *i.e.*, equivalent to 1000 IU (international units) of vitamin D taken orally.

This amount of energy is calculated over the erythemal irradiance and not over the spectral irradiance [10].

There is a significant correlation between the minimal exposure time to the Sun for the production of previtamin $D_3$ and the maximum exposure time for the production of erythema. This exposure time range gives us a time interval in which sunbathing is healthy without developing erythema, providing enough energy for the vitamin D synthesis. This time interval was studied by Salum *et al*. [11] for Concepción del Uruguay (Argentina) in a whole year but only considered the erythemal irradiance. In that study the minimum time interval obtained for the minimum dose of vitamin D was 6.6 – 20.7 min.

This paper uses the synthesis of previtamin $D_3$ for the phototype "III" skin, according to the classification by Fitzpatrick [6]. We apply this to residents of Urcuquí (0.51° N, 78.2° W) in Ecuador, which is at an altitude of 2384 meters above sea level therefore the incidence of solar radiation is much higher than elsewhere in the country.

## 2 Materials and methods

For a better understanding, it is necessary to review some methodological concepts used in this paper. First, the way to represent the response of each tissue for the wavelength of radiation is through the *action spectrum*. The action spectrum describes the relative effectiveness of the biological response for different wavelengths [3].

Among all action spectra of biological effects, the most popular is the CIE erythemal action spectrum (Commission Internationale de l'Éclairage) because erythema is the most common effect on the population. This erythemal irradiance is obtained by integrating the spectral irradiance weighted by CIE reference action spectrum up to 400 nm and normalized at 297 nm.

In this paper we need to use the action spectra of vitamin D [12] and erythema [13]. In Figure 1, these two action spectra are presented. It is possible to see that the highest sensibility for vitamin D synthesis is at 298 nm (value equal to 1) and it has a range of response from 250 to 330 nm. In the case of the erythema production, its total response is between 285 and 300 nm.

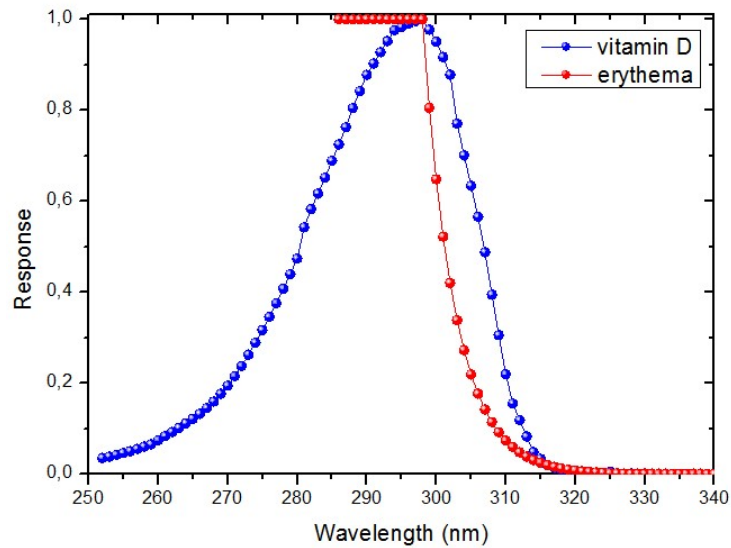

**Fig. 1.** Vitamin D action spectrum [12] (blue line) and erythema action spectrum [13] (red line).

Second, the intensity of solar radiation for each wavelength that reaches the Earth's surface is called *solar spectral irradiance*, (units of $W/(m^2\ nm)$). The solar spectral irradiance on March 15$^{th}$, 2015 can be seen in Figure 2 (top).

The determination of the solar irradiation from models is very useful when there is no equipment for measuring it. In the present article, the SMARTS model [14] allows us to obtain the solar intensity (irradiance) for each wavelength over the terrestrial surface in any instant of time (of day or year). This model requires input data of the weather conditions and the atmospheric constituents. This information was obtained from satellite databases, like: SSE/NASA (Solar meteorology and Solar Energy) which has solar and meteorological data; ESRL/NOAA (Earth System Research Laboratory of National Oceanic and Atmospheric Administration) about the average concentration (per month) of atmospheric $CO_2$; and

Giovanni/NASA (Goddard Earth Sciences Data and Information Services Center) which has a website with access to different satellites. The ozone data have been obtained from OMI/NASA and the aerosol data from MODIS-Aqua Deep Blue.

When the *spectral irradiance of biological effects* is needed, the mathematical convolution (the summation of the products for each wavelength) between the solar spectral irradiance I(λ,t) with the action spectrum ε(λ) of the biological effect, must be calculated. We will study the vitamin D ($I_{vitD}$) and the erythemal ($I_{ery}$) irradiances. The mathematical equations used to obtain each irradiance's biological effect are shown in the equations (1) and (2).

$$I_{ery}(\lambda, t) = \sum_{\lambda=286\,nm}^{400\,nm} I(\lambda, t) \cdot \varepsilon_{ery}(\lambda) \qquad (1)$$

$$I_{vitD}(\lambda, t) = \sum_{\lambda=252\,nm}^{330\,nm} I(\lambda, t) \cdot \varepsilon_{vitD}(\lambda) \qquad (2)$$

As can be seen from the equations (1) and (2), the erythemal action spectrum extends from 250 to 400 nm [15] and the spectrum of the production of pre-vitamin $D_3$ from 252 to the 330 nm [12]. Now, with $I_{ery}(\lambda,t)$ and $I_{vitD}(\lambda,t)$, the irradiance of biological effect can be obtained (in W/m$^2$) by integrating the spectral irradiance with regard to wavelength. The integral is a mathematical operation similar to the calculation of the area under a curve (eqs. 3 y 4). This process can be seen in Figure 2.

$$I_{vitD}(t) = \int_{\lambda_1}^{\lambda_2} I_{vitD}(\lambda, t)\, d\lambda \qquad (3)$$

$$I_{ery}(t) = \int_{\lambda_1}^{\lambda_2} I_{ery}(\lambda, t)\, d\lambda \qquad (4)$$

Each calculation of spectral irradiance and the following integration calculates only one point. In the present case, six points were calculated: one for every hour from 5:00 to 12:00 (local time). Since the solar irradiance is like a Gaussian curve, it must be symmetrical with regard to the solar noon. Then, the same six points calculated, until the solar noon, were used for 12:00 to 19:00 (local time). Finally, we did an interpolation of all points, in order to obtain the curve of the irradiance of the biological effect (Figure 2 in the right). In Figure 2, the arrow shows the correspondence between the area (upper panel) and the point on the curve (lower panel).

## 3 Results

Following the above-mentioned considerations, the interpolation was carried out using a Gaussian distribution (*cf*. eq. (5)) with parameters selected for each case. These ones are shown in Table 1.

$$I_{ery}(t) = y_0 + A \cdot e^{-\frac{1}{2}\left(\frac{t-t_c}{w}\right)^2} \qquad (5)$$

**Table 1.** Selected parameters for each day and each biological effect for the Gaussian distribution.

| Parameter | March 15th, 2015 | | December 15th, 2015 | |
|---|---|---|---|---|
| | Erythemal irradiance | Vitamin D irradiance | Erythemal irradiance | Vitamin D irradiance |
| $y_0$ | 0.001 | 0.001 | 0.001 | 0.001 |
| A | 0.403 | 0.830 | 0.348 | 0.677 |
| $t_c$ | 12:20 | 12:20 | 12:20 | 12:20 |
| w | 0.087 | 0.088 | 0.095 | 0.090 |

Because of the variability of the solar irradiance, due to the atmospheric parameters, each irradiance of biological effect will be different. We hope to later automate this calculation using software with routines calculating areas under curves.

The calculated vitamin D and erythemal irradiances from March 15th and December 15th, 2015, are presented in Figure 3 (left).

Once the irradiances of biological effect for erythema and vitamin D are calculated, we proceeded to calculate the needed exposure times to obtain the ¼ MED for phototype "III" skin. According to Fitzpatrick [6], the MED for UVB radiation is between 30-50 mJ/cm$^2$ for phototype "III". This is about 400 J (considering a medium value), so the ¼ MED is 100 J.

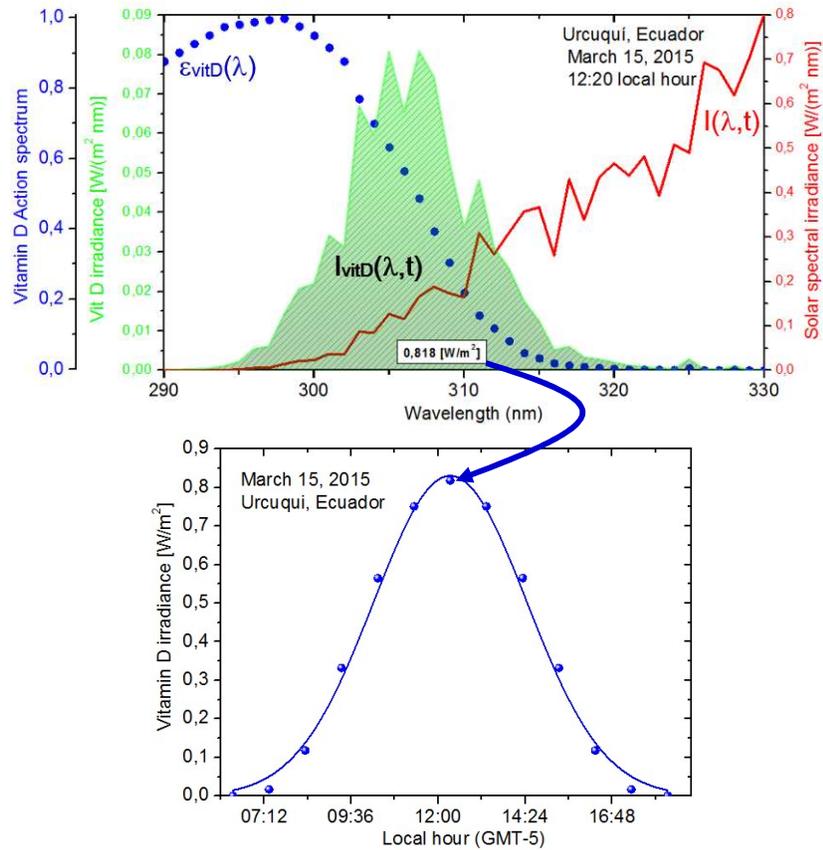

**Fig. 2. Top:** Solar spectral irradiance (red line), Vitamin D action spectrum (blue points) and Vitamin D spectral irradiance (green line). Vitamin D irradiance is shown as the shaded green area which is equal to the value of the integral. **Bottom:** Vitamin D irradiance calculated from the integrals (blue points) and a curve of Vitamin D irradiance using these points (blue line).

Table 2. Time of Sun exposure (min) that produces 100 J of energy.

| Local Time | March 15$^{th}$, 2015 | | | December 15$^{th}$, 2015 | | |
|---|---|---|---|---|---|---|
| | *Vitamin D* | *erythema* | *% Dif* | *Vitamin D* | *erythema* | *% Dif* |
| 5:00 | **125** | 160 | 28.0 | **124** | 142 | 14.5 |
| 6:00 | **73** | 104 | 42.5 | **73** | 89 | 21.9 |
| 7:00 | **34** | 58 | 70.6 | **36** | 49 | 36.1 |
| 8:00 | **15** | 28 | 86.7 | **17** | 25 | 47.1 |
| 9:00 | **7** | 14 | 100.0 | **8** | 13 | 62.5 |
| 10:00 | **4** | 8 | 100.0 | **5** | 8 | 60.0 |
| 11:00 | **3** | 5 | 66.7 | **3** | 6 | 100.0 |
| 12:00 | **3** | 5 | 66.7 | **3** | 5 | 66.7 |

We calculated the minimum exposure time for each hour, starting at 5:00, until 12:00 local time. For this, the integral of the irradiance was calculated to reach the condition of 100J. The minimum exposure time for each analyzed day and irradiance is presented in Table 2 and Figure 3 (right).

In Table 2, you can see the difference (in percentage) between the calculations made with the two irradiances (vitamin D and erythema). In some cases, the difference is 100% (*cf.* at 09:00 and 10:00 hours on March 15[th], 2015).

If other phototypes are considered for study, the MED for each one can be consulted on the scale proposed by Fitzpatrick. After an inspection of this scale it is easy to see that lower phototype means lower MED, and vice-versa. For example, a skin of phototype "I" can receive a MED of UVB radiation of 15-30 mJ/cm$^2$, on the other hand, a phototype "VI" can receive 90-150 mJ/m$^2$ before erythema emergence. To sum it up, if we want a model for all phototypes, we need to extend the present calculation to other skin types [16].

Figure 3 (right) depicts the minimum exposure time for obtain an energy of 100 J, using the erythemal (red) and the vitamin D (blue) irradiances on two different days (*i.e.*, March 15[th], 2015 –solid circles- and December 15[th], 2015 –empty circles-). According to Figure 3 (right), there is an important difference between the erythemal or vitamin D irradiances for both days, namely, the former irradiance predicts higher exposure times.

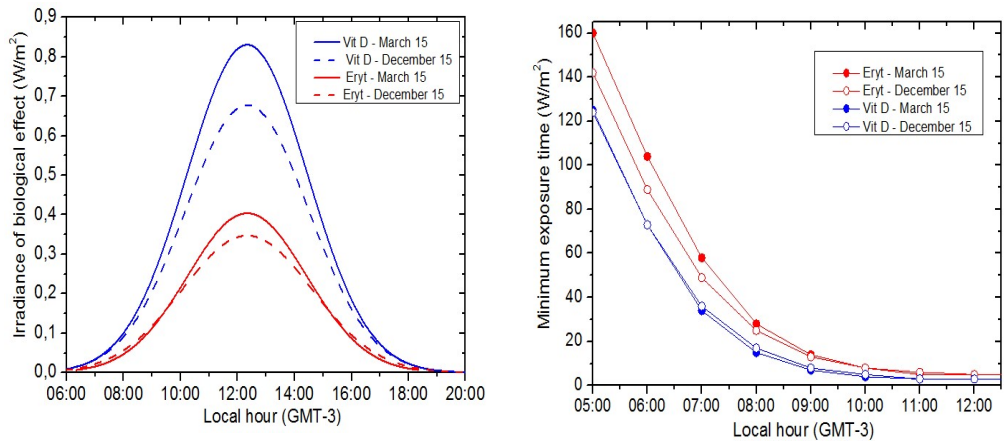

**Fig. 3.** Left: Vitamin D and erythemal irradiances for the two days **analyzed**. Right: Minimum exposure times using vitamin D and erythemal irradiances.

Holick's rule for Vitamin D is based on the MED concept related to the erythemal effect, but the spectral action of this effect is different from the one for the production of vitamin D. When Table 2 is analyzed it is clear that the minimum exposure time for the synthesis of vitamin D has greater values when the erythemal spectrum is used. Therefore, we can conclude that the calculation of the exposure time with the erythemal model produces larger minimum time values.

## 4 Conclusions

In this paper the estimation of the Sun exposure time in order to reach the minimum energy for the synthesis of vitamin D was studied. For comparison purposes, the solar spectral irradiance was calculated for two different days. The effect of the atmospheric constituents was considered consulting several satellite databases.

The minimum exposure times using the erythemal irradiance were higher than those using the vitamin D irradiance.

We will next ivestigate phototype "IV", which is very common in Ecuador. Furthermore, it is important to compare these results with other regions around the world with different values of latitude and altitude than the ones of Urcuquí. Finally, another interesting research line will be the effect of creams for different solar factors if we modify the minimum exposure time.

## Acknowledge

The authors wish to thank Professor Elisabeth Griewanck for their selfless contributions to our work.